\documentclass[pre,aps,twocolumn,superscriptaddress,floatfix]{revtex4-1}
\usepackage[english]{babel}

\usepackage{amsmath} 
\usepackage{color}
\usepackage{graphicx}% Include figure files
\usepackage{dcolumn}% Align table columns on decimal point
\usepackage{bm}% bold math
\newcommand{\betar}{\beta_{\mathbf{r}}}

\begin{document}

\preprint{APS/123-QED}

\title{Infection fronts in randomly varying %quenched 
transmission-rate media}% Force line 

\author{Renzo Zagarra}
\email{renzo.zagarra@ib.edu.ar}
\affiliation{Centro At\'omico Bariloche and Instituto Balseiro,
CNEA,  Universidad Nacional de Cuyo, 8400 Bariloche, Argentina}
%\homepage[]{Your web page}
%\thanks{}
%\altaffiliation{}
%\affiliation{}

\author{Karina Laneri}
\email{karina.laneri@ib.edu.ar}
\affiliation{Centro At\'omico Bariloche and Instituto Balseiro,
CNEA, CONICET and Universidad Nacional de Cuyo, 8400 Bariloche, Argentina}
%\homepage[]{Your web page}
%\thanks{}
%\altaffiliation{}
% \affiliation{}
\author{Alejandro B. Kolton}
\email{alejandro.kolton@ib.edu.ar}
\affiliation{Centro At\'omico Bariloche and Instituto Balseiro,
CNEA, CONICET and Universidad Nacional de Cuyo, 8400 Bariloche, Argentina}

\date{\today}

\begin{abstract}
We numerically investigate the geometry and transport properties of infection fronts within the spatial SIR model in two dimensions. The model incorporates short-range correlated quenched random transmission rates. Our findings reveal that the critical average transmission rate for the steady-state propagation of the infection is overestimated by the naive mean-field homogenization. Furthermore, we observe that the velocity, profile, and harmfulness of the fronts, given a specific average transmission, are sensitive to the details of randomness. In particular, we find that the harmfulness of the front is larger the more uniform the transmission-rate is, suggesting potential optimization in vaccination strategies under constraints like fixed average-transmission-rates or limited vaccine resources.
The large-scale geometry of the advancing fronts presents nevertheless robust universal features and, for a statistically isotropic and short-range correlated disorder, we get 
a roughness exponent $\alpha\approx 0.42 \pm 0.10$ and a dynamical exponent $z\approx 1.6 \pm 0.10$, which are roughly compatible with the one-dimensional Kardar-Parisi-Zhang (KPZ) universality class. We find that the KPZ term and the disorder-induced effective noise are present and have a kinematic origin. 
\end{abstract}

\maketitle
%\tableofcontents

\section{Introduction}

The general study of propagating fronts in random media is pertinent for modeling various out of equilibrium phenomena in physics, chemistry, biology and ecology~\cite{halpin1995,BarabasiBook,Krug1997,meakin1998,MurrayIIBook}. Examples encompass driven domain walls in ferromagnetic 
~\cite{Ferre2013,Ferrero2021,Wiese_2022} 
and ferroelectric ~\cite{Kleemann2007,Paruch2013} materials, imbibition fronts \cite{Planet2009}, chemical reaction waves
~\cite{kapral2012chemical,Chevalier2017},  epidemic waves ~\cite{Murray86,RohaniBook}, growth of bacterial colonies \cite{Allen_2019,Bhattacharjee2022}, cell fronts \cite{DeNardis2017},
two-species invasion~\cite{Malley2006}, and flame fronts~\cite{Provatas1995,Lam2017} among others. 
Considering such diverse microscopic complexities, a comprehensive investigation of propagating fronts would be formidable if not for the emergence of generic scale invariance and universal properties at sufficiently large spatial and temporal scales~\cite{kardar2007}.

From a statistical physics perspective, the most straightforward non-trivial portrayal of a rugged, condensed front advancing within a random medium, is that of a mobile interface. This significant model simplification, which conceals the specific local configuration of the front to emphasize its extensive long-wavelength transverse spatial fluctuations, facilitates the development of (theoretically convenient) minimalist continuum stochastic models exhibiting specific symmetries. These symmetries, in turn, give rise to distinct universality classes. These models serve as valuable tools for predicting the universal properties, both in transient and steady-states, of mobile interfaces. These properties can be quantitatively assessed against experiments or numerical simulations conducted using other models. Among these minimalist models, the paradigmatic Kardar-Parisi-Zhang (KPZ) equation~\cite{Kardar1986} has experienced a resurgence in interest owing to relatively recent theoretical advancements and experimental accomplishments~\cite{Halpin-Healy2015,sasamoto2016,takeuchi2018}.

On the one hand, a notable association between the KPZ equation and random matrices has revealed a more comprehensive universality, particularly regarding the precise height distribution of the KPZ interface~\cite{Quastel2015,corwin2012kardar}. On the other hand, the KPZ equation has demonstrated its accuracy in describing fronts across various experimental systems, including fronts observed in liquid crystals~\cite{Takeuchi2010} and reaction fronts~\cite{Atis2015}.
Experiments conducted on thin film turbulent liquid crystals, in particular, have enabled the precise determination of critical exponents and the universal height distribution with unprecedented accuracy~\cite{Takeuchi2010,DeNardis2017}.
The outstanding agreement with the KPZ predictions showcases the predictive capability of minimalist interface models.
Beyond these recent breakthroughs, the KPZ universality class has long been recognized for its breadth~\cite{BarabasiBook}, encompassing the geometric evolution observed in various established discrete models like ballistic deposition, Eden cluster growth, and solid-on-solid models, and also ecological invasion models \cite{OMalley2009}. Additionally, diverse mappings have enabled connections between the KPZ equation and models describing non-equilibrium transport properties of interacting particles, stirred fluids, as well as the equilibrium properties of a polymer in a random environment~\cite{halpin1995,BarabasiBook,sasamoto2016}.

In this paper, we tackle the issue of propagating fronts within the spatial Susceptible-Infected-Recovered (SIR) model in a random medium, as discussed recently in Ref.~\cite{Kolton2019}.
The interest in this specific model arises for three main reasons. Firstly, the spatial SIR model serves as a paradigmatic illustration of a two-component Reaction-Diffusion system, resembling an auto-catalytic chemical reaction. It generates a propagating infected front capable of irreversibly advancing over the susceptible population under specific conditions. Consequently, findings derived from the spatial SIR model can potentially serve as a representative model for comprehending more intricate reaction-diffusion systems.
Conversely, the spatial SIR model holds intrinsic interest and has found applications in modeling the spread of diseases such as rabies~\cite{Murray86} and other infection waves involving local transmission dynamics~\cite{RohaniBook,MurrayIIBook}. Additionally, with minor adjustments to the reaction term, the model has been extended to describe phenomena such as flame fronts~\cite{Provatas1995}. Moreover, while numerous precise outcomes have been achieved in the homogeneous scenario for the spatial SIR model~\cite{MurrayIIBook}, comparatively less is understood about the heterogeneous case in general.
Disorder, particularly when one or more parameters of the model are fixed but exhibit random spatial variations following well-defined distributions and correlations, can be significant in numerous systems. In this regard, recent numerical findings have demonstrated that a quenched random dichotomous transmission rate induces roughness in the front and significantly alters its velocity, amplitude, and shape compared to the homogeneous scenario.
A critical behaviour near the onset of steady-state propagation was also reported~\cite{Kolton2019}.

Here, we expand upon the findings presented in Ref.~\cite{Kolton2019} in both quantitative and qualitative aspects. We explore various forms of isotropic short-range correlated disorder and demonstrate that, even with a fixed spatial average of the transmission rate, the velocity and epidemiological impact of the front significantly rely on the specifics of the disorder. Additionally, our observations indicate that the kinetic roughening of an infection front remains robust despite these variations and roughly aligns with the characteristics of the one-dimensional KPZ universality class. This behavior involves a kinetically generated KPZ term alongside noise induced by quenched disorder.

\section{Model, Method and Properties}

We model the spread of an infectious disease through a two-dimensional heterogeneous space. We consider a local fraction of susceptible individuals $S(\mathbf{r}, t)$ and a fraction of infected individuals $I(\mathbf{r}, t)$. The susceptible individuals are assumed to be immobile, while the infected individuals are assumed to diffuse with a diffusion constant $D$. The susceptible fraction at the position $\mathbf{r}$ can be converted into infected by local contact with the infected fraction, with a position dependent transmission rate $\beta_{\mathbf{r}}$. Infected individuals recover at a constant rate $\gamma$ and become part of the $R(\mathbf{r}, t)=1-S(\mathbf{r}, t)-I(\mathbf{r}, t)$ fraction. Under these assumptions, the dynamics of $S$ and $I$ can be described by the well known diffusive SIR model, \cite{MurrayIIBook,RohaniBook}:
\begin{align}
    \partial_t S & = -\beta_{\mathbf{r}}SI, \label{eq:Seq}\\
    \partial_t I & = \beta_{\mathbf{r}} SI - \gamma I + D \nabla^2 I.
    \label{eq:Ieq}
\end{align}
with the recovered fraction not playing any role in the dynamics.

\begin{figure}[h]
\centering
\includegraphics[width=.49\textwidth]{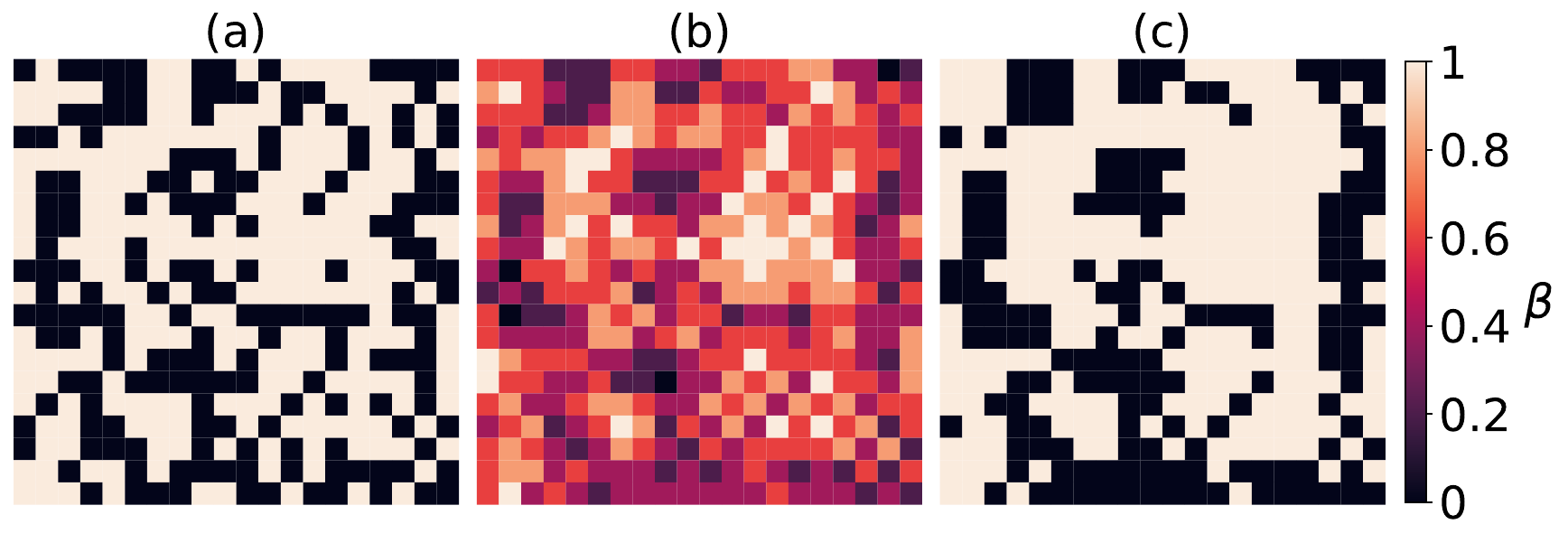}
\caption{Zoomed snapshots of typical realizations of the three types of spatially random 
transmission rates considered in this work: 
(a) Random dichotomous heterogeneity (RD); 
(b) Smoothed random dichotomous heterogeneity (SRD); 
(c) Correlated dichotomous heterogeneity (CD). All shown cases correspond to $p=0.5$.}    \label{fig:betas}
\end{figure}

To detect universal and non-universal transport properties of the infection fronts we will consider four different statistically homogeneous types of heterogeneous random transmission rates $\beta_{\mathbf{r}}$, that we will call H, RD, SRD and CD:  
\begin{itemize}
    \item H: $\betar=\beta$ is the homogeneous case. This is a well known case \cite{MurrayIIBook} we use just for reference.
    \item RD: 
 $\betar$ is a random dichotomous spatially uncorrelated  heterogeneity described by the Bernoulli distribution \cite{Kolton2019}
\begin{equation}
    f(\beta_{\mathbf{r}}) = p\delta (\betar) + (1-p)\delta(\betar - \beta),
\end{equation}
with $0 \leq p \leq 1$. Conceptually, $p$ measures the fraction of space where infection can not take place. In other words, we can also think $p$ as the fraction of the population that was ``vaccinated'' at randomly selected places where we set $\beta_{\bf r}=0$.

\item SRD: $\betar$ is a smoothed version of the previous case, where we average over the first  neighbors in a square discretized lattice in order to smooth the RD binarized heterogeneity. This case allows us to analyze the effect of both smoothing and of a slight increase of the correlation length with respect to the RD case.

\item CD: $\betar$, in this correlated dichotomous heterogeneity, comes from rebinarizing SRD after a single step using a threshold given by the mean spatial transmission rate of the original SRD.  In this case we slightly increase the range of correlations with respect to the RD case but retaining the binary distribution. This type of heterogeneity allows us to disentangle the effect of local correlations from the effect of smoothing, present in the SRD.
\end{itemize}
In all cases, disorder is parameterized by $\beta$ and $p$.  The variation of $p$ allows to vary the spatially averaged transmission rate $\overline{\beta}\equiv \overline{\beta_{\bf r}} $ which is a convenient control parameter as the local reduction of $\beta_{\bf r}$ in e.g., epidemics or forest fires, implies vaccination or protection applied locally. Fixing $\overline{\beta}$ is thus equivalent to compare cases with such protection resources fixed. 
In Figure \ref{fig:betas} we show typical zoomed snapshots of each type of disorder in a square-lattice discretized system, for $p=0.5$.

We will be interested in the evolution of an initially flat travelling infection front that emerges by introducing a flat initial infected fraction $I(\mathbf{r},t=0) = I_0\Theta(\delta_x-x)$, 
with $I_0=1$,
%with $I_0$ large enough to trigger a front
on a stripe of size $\delta_x$ over an uniform initial susceptible fraction $S(\mathbf{r},t=0)=S_0$. 
For the homogeneous case (H) for instance, if 
$S_0 > \gamma/\beta$
% $S_0 > S_c \equiv \gamma/\beta$ 
any $I_0>0$ will trigger a traveling wave, leaving behind a reduced fraction of susceptible $S_1 < \gamma/\beta < S_0$~\cite{MurrayIIBook,Kolton2019}.  

We will consider a two-dimensional space of size $L_{\rm x} \times L_{\rm y}$, such that $L_{\rm x}\gg \max{(\delta_x,\Delta})$, with $\Delta$ the characteristic intrinsic front width and $\delta_x$ the initial stripe of infected. We use Dirichlet boundary conditions in the x-direction, $I(x=0,y,t) = S(x=0,y,t) = I(x=L_{\rm x},y,t) = S(x=L_{\rm x},y,t)=0$ and periodic boundary conditions in the y-direction, $I(x,y=0,t) = I(x,y=L_{\rm y},t)$ and $S(x,y=L_{\rm y},t) = S(x,y=L_{\rm y},t)$. These initial and boundary conditions are one of the many possibilities to obtain a single front propagating in the positive $x$-direction which can get rough but remaining flat on average.

 During the simulations we monitor the evolution of $I(x,y,t)$ and compute different observables. The simulations stop when the most advanced point of the front reaches $x=L_{\rm x}$. During the corresponding time interval we can identify transient and steady-state properties of the propagation. Without loss of generality we have fixed $D=1$, $\beta=1$ and $\gamma=0.2$, $S_0=1$, in all the numerical simulations.

In order to characterize the infection front we define an univalued displacement field $u(y,t)$ from
\begin{equation}
  u(y,t) = \frac{\overline{x I(x,y)}}{\overline{I(x,y,t)}} \equiv \frac{\int_{x,y} x I(x,y,t)}{\int_{x,y} I(x,y,t)}, 
  \label{eq:campo}
\end{equation}
where $\overline{[...]}$ denotes average over the two dimensional space. The center of mass of the front is
\begin{equation}
  u_{\rm cm}(t)\equiv
  \int_y \frac{u(y,t)}{L_{\rm y}}.
  \label{eq:centromasa}
\end{equation}
We then define the disorder averaged
instantaneous center of mass front velocity as
\begin{equation}
  c(t) \equiv \langle \dot{u}_{\rm cm}(t) \rangle,\label{eq:velocidad}
\end{equation}
with $\langle \dots \rangle$ denoting average over disorder realizations.
To characterize spatial fluctuations we define  the {\it roughness},
\begin{equation}
  \omega(t)\equiv
  {
  \left\langle
  \int_y \frac{[u(y,t)-u_{\rm cm}(t)]^2}{L_{\rm y}} \right\rangle^{1/2}},
  \label{eq:rugosidad}
\end{equation} 
and the instantaneous structure factor
\begin{equation}
F(q,t) \equiv 
\langle
|u(q,t)|^2
\rangle
,\label{eq:Sfactor}
\end{equation}
where $u(q,t)$ is the Fourier transform of $u(y,t)$.

It's important to note that, in general, the front crest can exhibit a multi-valued nature concerning $y$, potentially involving overhangs and pinch-off loops. In such instances, it cannot be accurately described solely by $u(y,t)$ as defined in Eq.~\eqref{eq:campo}.
However, in the scenario we investigate, where periodic boundary conditions in the $y$ direction and an initial condition $u(y,t=0)=u_0$ are applied, $u(y,t)$ continues to serve as a suitable approximation for characterizing the front's velocity and geometry. Under these conditions, although the front becomes rough, super-roughening does not manifest for the types of disorders we have examined. Consequently, the large-scale geometry of the front can be accurately described by the single-valued function $u(y,t)$.
The purpose of this simplified portrayal of the front is to roughly equate the problem to that of interfaces advancing through random media and to uncover potential universal behaviors~\cite{BarabasiBook}.

To extract other relevant properties of the front that go beyond the reduction to a simple evolving interface we additionally define the average amplitude of the front as
\begin{equation}
  I_{\rm max}(t)=
  \left\langle
  \int_y  \frac{I[u(y,t),y,t]}{L_{\rm y}}\right\rangle,\label{eq:maximo}
\end{equation}
and the {\it noxiousness} of the front as the fraction of susceptible sites after the front have completely propagated through the medium
\begin{align}
    S_1 =   \left\langle
\overline{S(x,y,t=T)}\right\rangle, 
    \label{eq:S1}
\end{align}
where $T$ is the moment where the foremost point of the front touches $x=L_{\rm x}$. The lower this number, the higher the ``epidemiological harmfulness'' of the front.

\section{Results}

\subsection{Transport properties}

\begin{figure}[h]
   \includegraphics[width=0.5\textwidth]{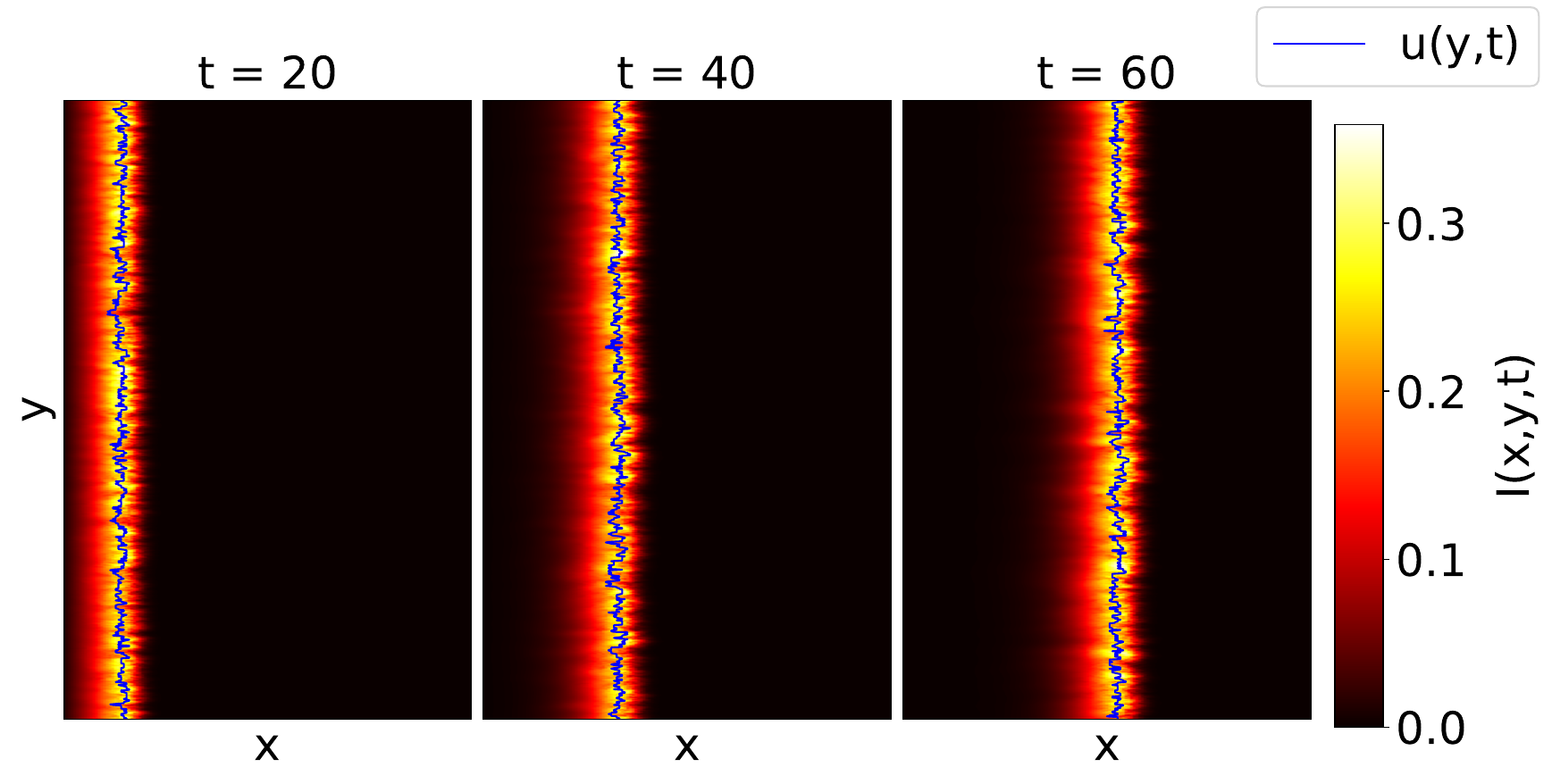}
    \caption{
    Example of an evolving infection front in a RD heterogeneous medium with $p = 0.3$, 
    for three times following an initially flat ignition at the left, $I(x,y,t=0)=I_0 \Theta(\delta_x-x)$. 
    The blue line shows the front displacement field $u(y,t)$, from Eq.\eqref{eq:campo}. The colormap shows the local and instantaneous infective amplitude $I(x,y,t)$. 
    }
    \label{fig:fronts}
\end{figure}

In Figure \ref{fig:fronts} 
we show snapshots of $I(x,y,t)$ at different times for the RD type of disorder. As observed in the homogeneous case the front keeps an asymmetric shape in presence of disorder, with a leading edge decaying faster than the trailing edge, but both decaying exponentially fast in a characteristic distance $\Delta \sim D/c$~\cite{Kolton2019}. The rough and fluctuating displacement field $u(y,t)$ (Eq. \eqref{eq:campo}) is also appreciated.

\begin{figure}[h]
    %\centering
    \includegraphics[width=0.5\textwidth]{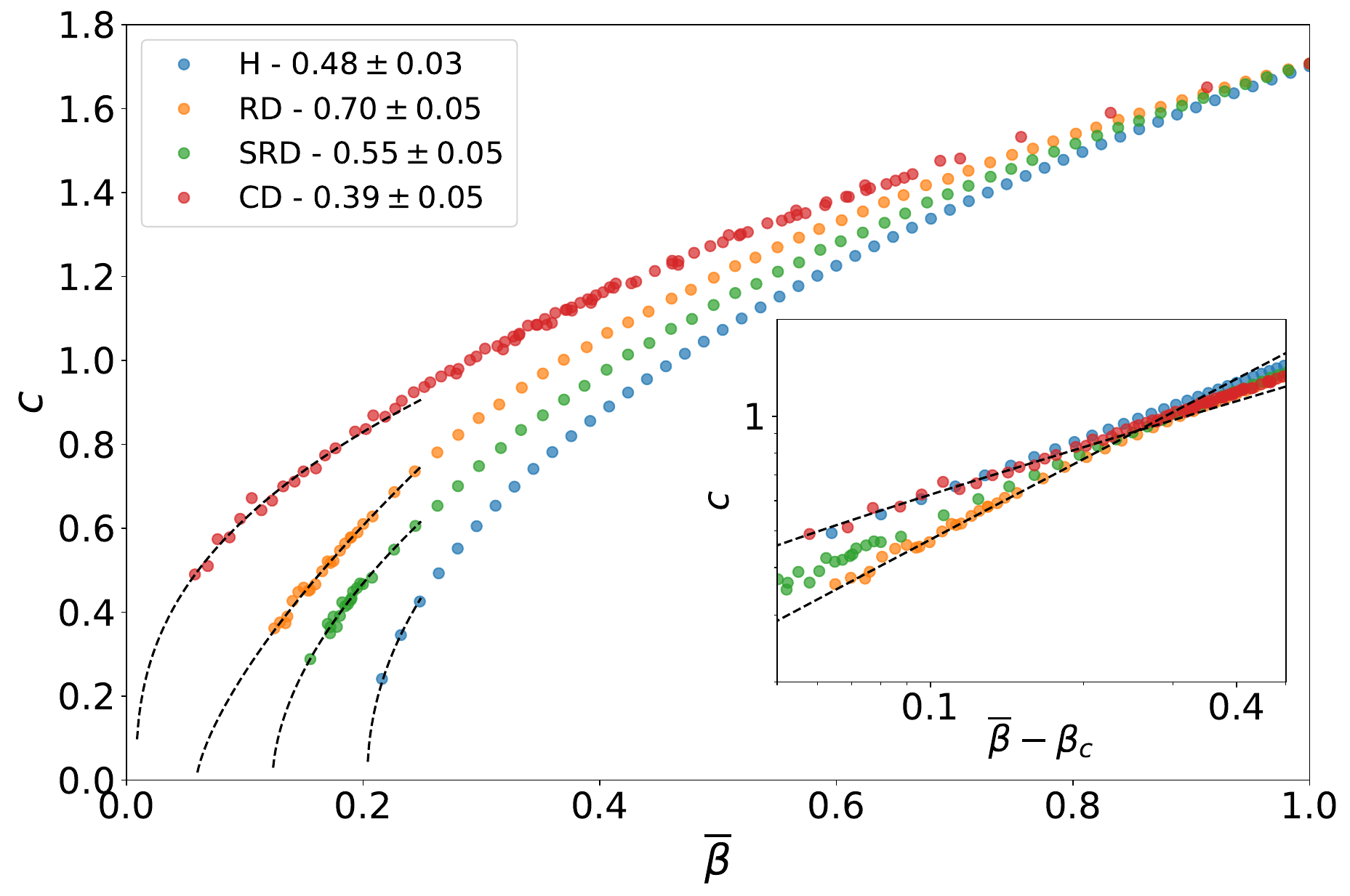}
    \caption{
    Infection front velocity $c$ as a function of the spatially averaged transmission-rate $\overline{\beta}$ 
    in different media, as indicated. Dash-lines are power-law fits near the infection threshold. 
    The inset highlights the critical behaviour near the infection thresholds corresponding to each type of media. 
    }
    \label{fig:velocities}
\end{figure}

We find that after a microscopic transient the front acquires a well defined steady-state mean velocity $c=\langle c(t) \rangle$, such that $u_{\rm cm}(t) \sim c t$ at long enough times. 
We have also checked that $c$ is size independent above a finite value of $L_y$.
In Figure \ref{fig:velocities} we show $c$ vs the spatially averaged quenched transmission rate $\overline{\beta}$
for different types of disorders.  
Note that for the RD case $\overline{\beta}=\beta(1-p)$ while for the other cases are more complicated functions of $p$ and $\beta$, but we can always fix $\overline{\beta}$ for all the considered cases.
The homogeneous (H) case is shown for reference. As we can appreciate, in all cases there is a critical value for $\overline{\beta}$, below which the front does not propagate, and above which a propagating front exists with a velocity $c$ that increases monotonically with $\overline{\beta}$. We also observe that near the threshold for propagation $\beta_c$ there is an approximate power-law behaviour $c\sim (\overline{\beta}-\beta_c)^\alpha$ indicating a critical phenomenon (see inset) with a critical exponent $\alpha<1$ in all cases and disorder dependent thresholds, 
that we will thus denote $\beta_c$ by 
${\beta}_H$, ${\beta}_{RD}$, ${\beta}_{SRD}$ and ${\beta}_{CD}$, for each type of disorder. 
It is worth noting that $\alpha<1$ exponents are similar to those observed for an elastic interface depinning transition in random media, for the mean velocity as a function of the driving force~\cite{Ferrero2021}.
Note however that at variance with the interface depinning transition, infection fronts broaden and vanish as we approach the thresholds, and can not persist if they are static. The critical configuration of the infection front is thus not well defined ~\cite{Kolton2019}.

\begin{figure}[h]
    %\centering
\includegraphics[width=0.5\textwidth]{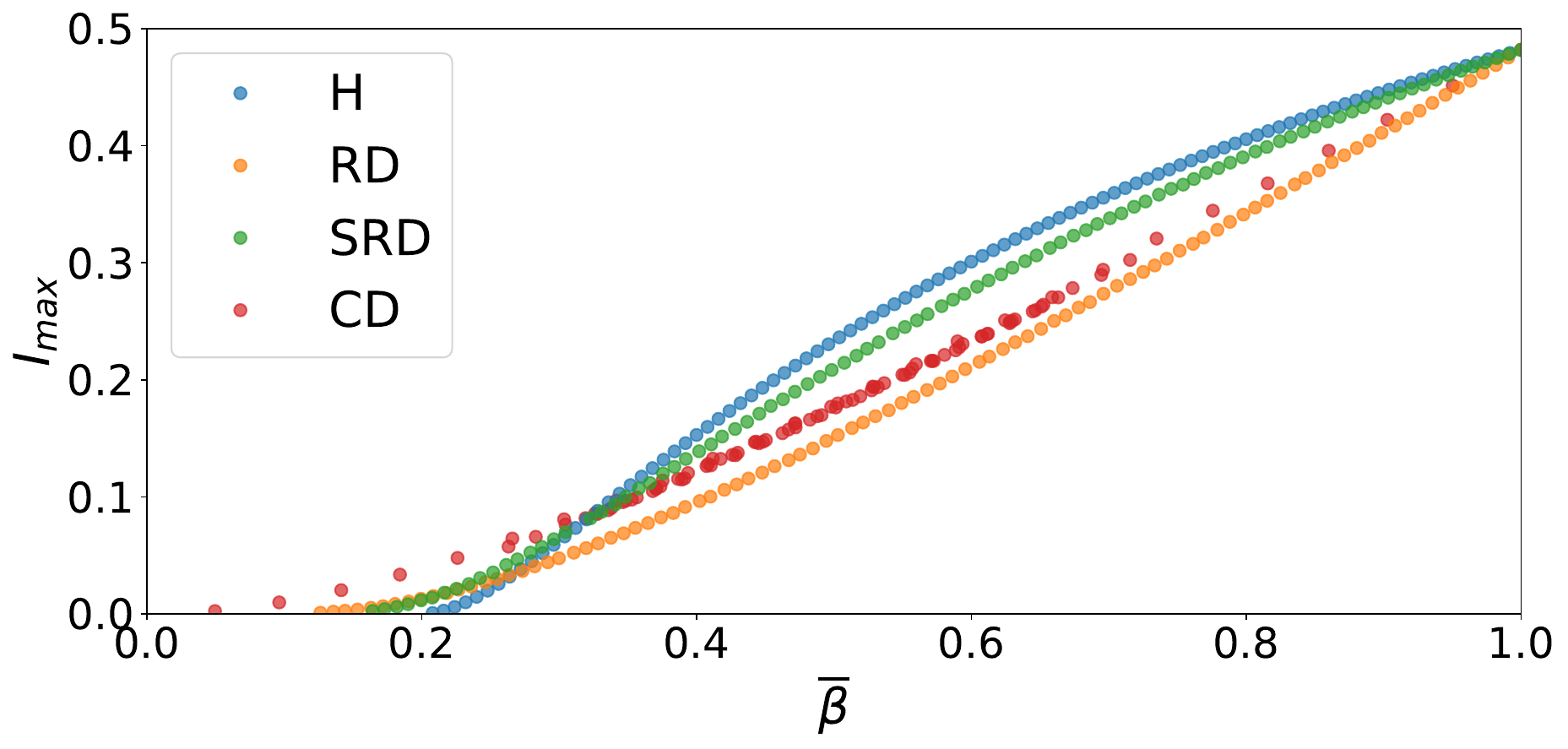}
    \caption{
    Infection front amplitude $I_{\rm max}$ as a function of the spatially averaged transmission-rate 
    $\overline{\beta}$ in different media, as indicated. The thresholds for different media 
    correspond to the velocity thresholds of Figure \ref{fig:velocities}.  
    }
    \label{fig:amplitude}
\end{figure}

The critical behaviour of Figure~\ref{fig:velocities} is qualitatively expected. For the H case we know for instance that the threshold for propagation is $\beta_{\rm H}=\gamma/S_0$, the same condition that applies in the mean-field SIR model. The velocity above the threshold in this case is predicted to be 
\begin{align}
c_{\rm H} = 
%2\sqrt{D(\beta S_0-\gamma}= 
2\sqrt{D S_0(\overline{\beta}-\beta_{\rm H})}, 
%\sim (\beta-\beta_c)^{1/2}  
\label{eq:cH}
\end{align}
which agrees well with the fitted critical region, yielding $c_{\rm H} \sim (\overline{\beta}-\beta_{\rm H})^{\alpha_{\rm H}}$ with $\alpha_{\rm H} = 0.48 \pm 0.03$.
A naive mean-field approximation for the non-homogenous cases predicts the same result obtained for $c_{\rm H}$ for $c_{\rm RD}, c_{\rm SRD}, c_{\rm CD}$ cases, as a function of $\overline{\beta}$. In contrast, in Figure \ref{fig:velocities} we observe  a different velocity vs $\overline{\beta}$ curve for each case. We first observe that the thresholds for propagation satisfy 
$\beta_{\rm CD}<\beta_{\rm RD}<\beta_{\rm SRD}<\beta_{\rm H}$. It is therefore easier to ignite a travelling front by increasing the average transmission rate $\overline{\beta}$ in the disordered cases.
Among the disordered cases we also observe that the smoothed disorder has the largest threshold.
We also observe that the critical exponent differ from the $\alpha_{\rm H} \approx 1/2$ value (see inset), although it is difficult to get an accurate value of the exponents because fronts do not only have a vanishing velocity but also a vanishing amplitude (see Figure \ref{fig:amplitude}) and a diverging intrinsic width $\Delta$. Interestingly, for a fixed value of $\overline{\beta}>\beta_{\rm H}$, the velocity follows the order $c_{\rm CD}>c_{\rm RD}>c_{\rm SRD}>c_{\rm H}$, with the minimum velocity corresponding to the H case and the maximum to the CD case.

As for the mean velocity $c$,  we find that after a microscopic transient the front acquires a well defined steady-state amplitude $I_{\rm max}$. Moreover, it was shown that during such transient, the $x$-profile function of the front $\int_y \langle I(x,y,t)\rangle/L_y $ reaches a steady-state \cite{Kolton2019}.
We have also checked that $I_{\rm max}$ is size independent above a finite value of $L_y$. 
In Figure \ref{fig:amplitude} 
we see that the steady-state amplitude of the fronts also follows a critical behaviour near the corresponding thresholds,
$I_{\rm max}\sim (\overline{\beta}-{\beta}_{c})^\theta$ with $\theta>1$, in all cases. 
At variance with $c$ however we can observe crossings of $I_{\rm max}$ at $\overline{\beta}\approx 0.3$. 
Above it, we can see that for a fixed $\overline{\beta}$ the homogeneous H case, which has also the smallest velocity, 
has the highest amplitude, while the RD, which has the larger velocity, has the lowest mean amplitude. It is worth pointing here that the condition $\overline{\beta}>\beta_c$ ensures a propagating front with a well defined mean velocity and amplitude, determined only by the bulk properties $\beta_{\bf r}, S_0, \gamma, D$, regardless of the amplitude $I_0$ (here chosen as $I_0=1$ only to minimize transients) of the initial infected stripe or boundary conditions in the $x$-direction.

Several of the observed differences between the homogeneous and heterogeneous cases can be elucidated by the existence of an array of preferred paths for front propagation, akin to the movement of an invading fluid in porous media.
These preferred paths, characterized by a higher concentration of transmission rate, suggest a swifter front velocity in the heterogeneous cases compared to the homogeneous case. 
Conversely, the threshold for propagation is expected to be lower in the heterogeneous case, allowing the initiation of the front in these favored paths even when the average transmission rate remains below $\beta_{\rm H}$. This explanation accounts for the two primary differences depicted in Figure \ref{fig:velocities}.
A caveat of this qualitative argument is that the preferred paths are not percolating paths. For instance, in the RD and CD cases, the probability of a percolating path connecting sites with $\beta_{\bf r}=\beta$ is negligible when $\overline{\beta}$ is small. However, sites or clusters of sites with $\beta_{\bf r}=\beta$ can still be connected by the diffusion of infected individuals, forming weak links between them. These weak links ultimately constitute preferred, albeit tortuous, paths for the propagation of the entire front.
Using a heuristic model to bolster these qualitative arguments, let's consider a scenario where the two-dimensional space is partitioned into $2M$ equal homogeneous stripes oriented along the $x$ direction. These stripes alternate along the $y$ axis between $\beta_{\bf r}=2\beta$ and $\beta_{\bf r}=0$, ensuring that the average $\overline{\beta}$ remains fixed at $\beta$. This setup, a rare representation of the CD case, would result in an extended front, approximately exhibiting a velocity higher than that in the homogeneous case H (as given by Eq.\eqref{eq:cH}), expressed as $c(\overline{\beta})\approx 2\sqrt{2 D S_0(\overline{\beta}-\beta_{\rm H}/2)}>c_{\rm H}(\overline{\beta})$. Furthermore, the propagation threshold would also be lower compared to the H case. These two predictions align with numerical simulations of this model using periodic stripes, particularly when $M$ is not considerably large, minimizing the impact of transverse infected diffusion and interconnection between stripes.
The heuristic model's predictions regarding velocity and threshold are qualitatively consistent with the outcomes depicted in Figure \ref{fig:velocities}. Additionally, it elucidates why the SRD exhibits a lower velocity than the RD and CD cases. In the SRD, the smoothing effect blurs the convenient paths present in the RD, making it closer to the H case. Moreover, it clarifies why $c_{\rm CD}>c_{\rm RD}$, as the CD case is anticipated to possess less tortuous and thicker paths than the RD.

\begin{figure}[h]
    %\centering
    \includegraphics[width=0.5\textwidth]{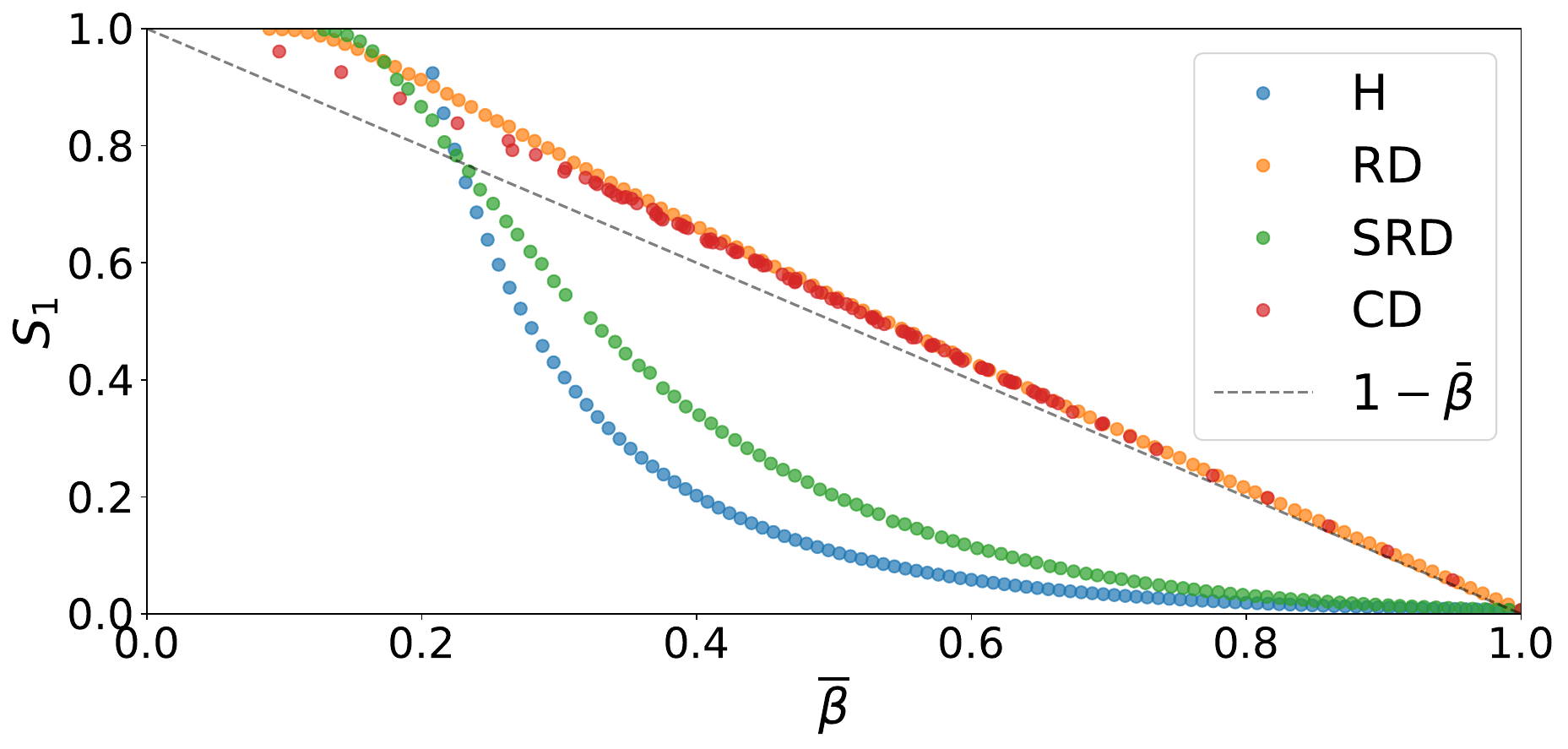}
    \caption{
    Remaining susceptible fraction 
    $S_1$ (Eq.\eqref{eq:S1}) after the passage of the infection front in different media, as indicated, 
    as a function of the spatially averaged transmission-rate $\overline{\beta}$. 
    The dashed-line indicates the expected assymptotic behaviour as $\overline{\beta}\to 1$ (see text).
    }
    \label{fig:nocividad}
\end{figure}

Finally, to quantify the damage caused by the front in the various media depicted, we display  in Figure \ref{fig:nocividad} the fraction $S_1$ of susceptibles remaining after the front's passage, as given by Eq.\eqref{eq:S1}, plotted as a function of $\overline{\beta}$.
For a fixed $\overline{\beta}>0.3$, the homogeneous case $H$ is the more harmful, followed in order by the SRD, the CD and the RD as the less noxious. 
Interestingly, Figure \ref{fig:nocividad} shows that the binarized transmission-rates yield a more effective protection than the smoothed cases, SRD or H, for the same $\overline{\beta}$. 
The CD and RD show a similar simple linear dependence that can be explained by noticing that a fraction $p$ of the sites are completely protected, and they dominate $S_1$. Therefore, $S_1 \sim p=1-\overline{\beta}/\beta$, explaining the dependence 
for high $\overline{\beta}$. 
It is also worth noting that in the homogeneous H case $S_1$ is expected to be larger the larger the threshold $\beta_{\rm H}=\gamma/S_0$~\cite{MurrayIIBook}. Our results for the heterogeneous cases for a fixed $\overline{\beta}>0.3$ actually show that the epidemic severity is in all cases lower than in the homogeneous H case in spite of having lower thresholds.

In summary, {\it for a fixed $\overline{\beta}$} or a fixed amount of vaccination used to locally suppress $\beta_{\bf r}<\beta$ from the basic and uniform transmission rate  $\beta$, we have found that:
\begin{itemize}
    \item The threshold for the front propagation is the largest in the homogeneous H case, followed by the SRD, the RD and the CD. 
    \item Fronts in the heterogeneous cases move faster and thus spend less time in the medium. 
    \item Fronts in the heterogeneous cases have a lower average amplitude. 
    \item After the propagation of the front, the total fraction of susceptible is higher in the heterogeneous cases, particularly in the dichotomous cases. 
\end{itemize}

From an epidemic perspective, these features can be advantageous or disadvantageous. Heterogeneous media are more susceptible to ignite from an initial presence of infection. However, the fronts in the heterogeneous case spread faster than in the homogeneous case and are generally less severe. This means that fewer individuals suffer from the passage of the infection front. Moreover, the observation that the average amplitude of the front is lower in heterogeneous cases appears as an advantage, especially if external actions can be applied to control an already propagating front, such as in managing a forest fire or controlling an epidemic.

\subsection{Kinetic Roughening}
We will now delve into the geometry of the interface that delineates the advancing front crest in different media. In the homogeneous (H) case, the front remains flat as it propagates, leading to a two-dimensional problem that simplifies into a one-dimensional one. However, in the presence of heterogeneity, symmetry is disrupted in all directions. At variance with Ref.\cite{Kolton2019} we focus here on the large-scale geometry at finite velocities, in order to avoid the critical broadening and weakening of the front expected at very low velocities near the propagation threshold corresponding to each type of disorder
~\cite{Kolton2019}.

As depicted in Figure~\ref{fig:fronts}, we observe that initially flat configurations of fronts become rough.
In contrast with velocity $c$ (Eq.~\eqref{eq:velocidad}) and the amplitude $I_{\rm max}$ (Eq.~\eqref{eq:maximo}) which reach steady values after a microscopic transient, the roughness $\omega(t)$, defined in Eq.~\eqref{eq:rugosidad} and the structure factor $F(q,t)$ (Eq.~\eqref{eq:Sfactor}),  achieve a steady-state saturation value 
$\omega_{sat}$ after a macroscopic time $t_\times$. 
This difference is related to the existence of a growing correlation length and to the observation that above a finite $L_y$,  $c$ and $I_{\rm max}$ become $L_y$-independent and are thus self-averaging, while $\omega_{sat}$ scales with $L_y$, as discussed below.
As we show next, both $t_{\times}$ and $\omega_{\rm sat}$ grow with $L_{\rm y}$ as expected for interface roughening processes, yielding universal properties of the infection fronts~\cite{BarabasiBook}.

\begin{figure}[h]
\centering
\includegraphics[width=\columnwidth]{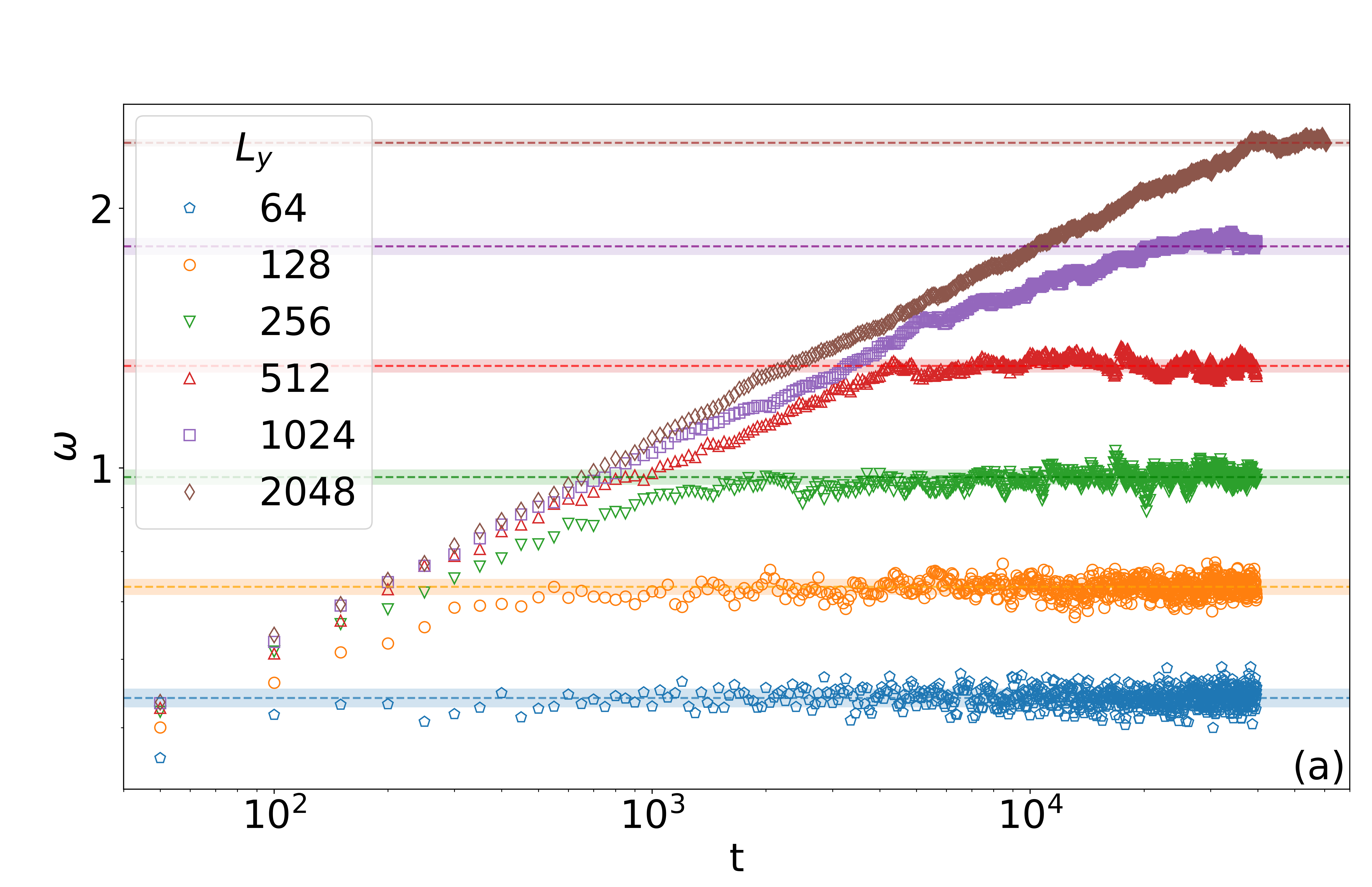}
      \includegraphics[width=\columnwidth]{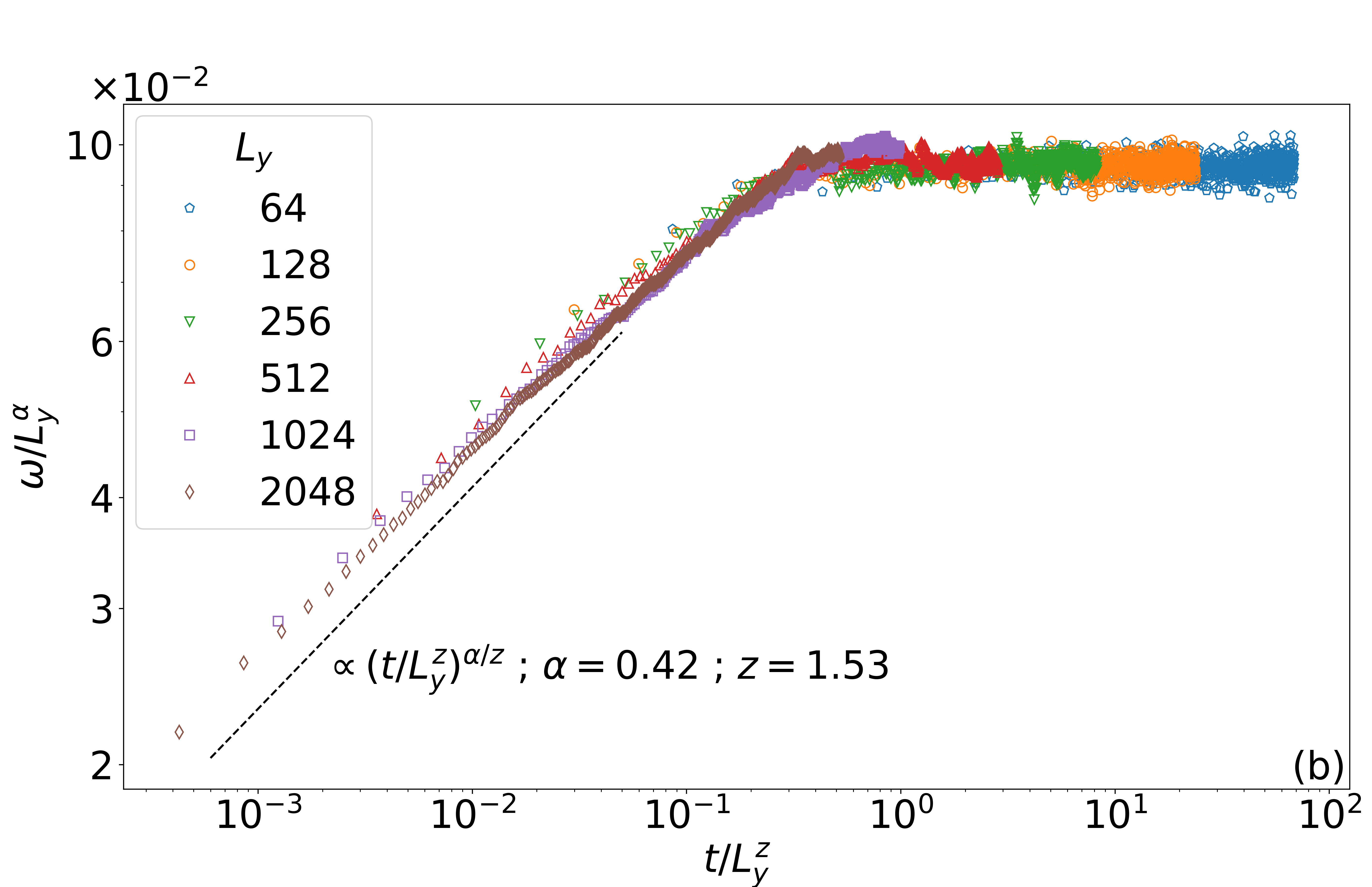}
\caption{
(a)
Front width $\omega$ (Eq. \eqref{eq:rugosidad}) vs time $t$ in the RD medium at a finite velocity $c \approx 1.7$, 
for different system sizes $L_{\rm y}$. We use $L_{\rm x}=2^{16}$ in order to ensure stationarity at large times. 
Saturation values (stationary $\omega$) for different $L_{\rm y}$ are marked by dashed-lines. 
Data was obtained after averaging over 150 disorder realizations of the disorder.
(b) Same curves rescaled, as indicated, into a master curve.
    }

    \label{fig:width_vs_t}
\end{figure}

In Figure \ref{fig:width_vs_t}(a) we show the mean square interface width $\omega(t)$ as a 
function of time $t$ for different system sizes $L_{\rm y}$,  for the RD case at a finite front 
velocity $c \approx 1.7$.
The scaling collapse of Figure \ref{fig:width_vs_t}(b) confirms a typical roughening process described by a Family–Vicsek type of scaling~\cite{BarabasiBook} with $\omega(t)\sim t^{\alpha/z}$ for $t < t_\times$ and a saturation $\omega(t)\sim L_{\rm y}^{\alpha}$ 
for $t > t_\times$, where $t_\times \sim L_{\rm y}^z$. In Figure \ref{fig:width_vs_t}(b) we see in particular that 
\begin{equation}
\omega = L_{\rm y}^{\alpha} f(t/L_{\rm y}^z),    
\label{eq:wfamilyviscek}
\end{equation}
with a master curve $f(x)\sim x^{\alpha/z}$ for small $x$ and $f(x)\sim~{\rm const}$ for large $x$, describes well the data.
Interestingly, similar curves are obtained for different finite velocities $c$ and also for the CD and SRD types of disorder. In all cases the exponents obtained from the fits are comprised in the range $\alpha=0.42 \pm 0.10$ and $z=1.6 \pm 0.10$ for the range of $L_{\rm y}$ we were able to access computationally. 
These exponents thus appear to be universal. At this respect we note that the roughness exponent  
$\alpha$ is nevertheless larger than the $\alpha=0.3 \pm 0.05$ obtained only for a RD disorder in Ref.\cite{Kolton2019} for interfaces of comparable sizes. As we discuss later, this difference might be attributed to the critical increase of the front intrinsic width near the infection threshold where the later exponent was estimated.

\begin{figure}[h]
    \centering
    \includegraphics[width=\columnwidth]{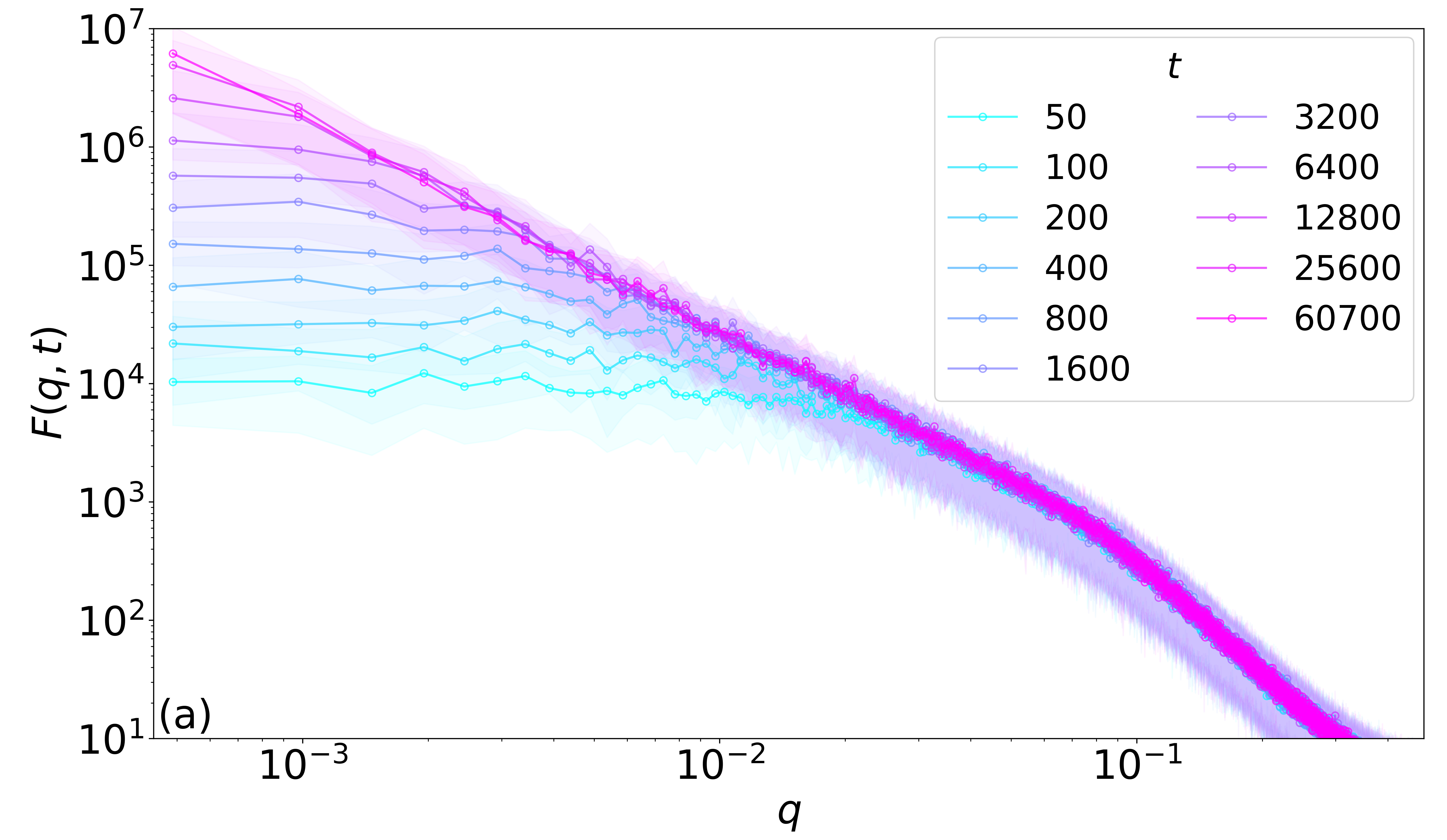}
    \includegraphics[width=\columnwidth]{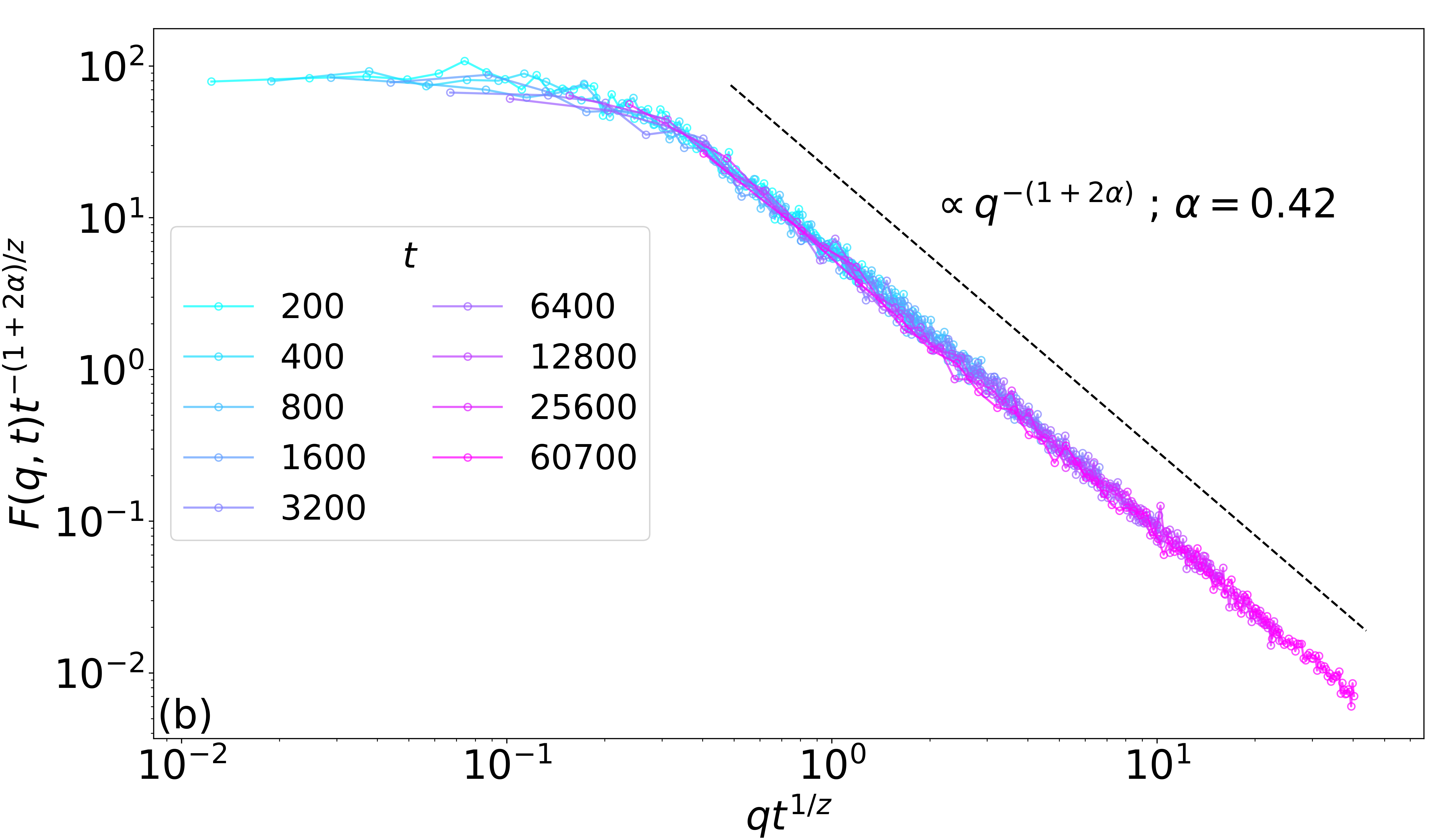}
    \caption{Structure factor $F(q, t)$ of the front vs time, averaged over 150 RD disorder realizations, for  
    a $L_{\rm y} \times L_{\rm x} = 2^{11} \times 2^{16}$ system. 
    \label{fig:Sq_vs_t}}
\end{figure}

The standard scaling and exponents just discussed are confirmed by the structure factor in Figure \ref{fig:Sq_vs_t}, with well defined power-law regimes. Figure \ref{fig:Sq_vs_t}(a) shows that the roughening process is controlled by a growing dynamical length scaling as $\sim t^{1/z}$ for $t<t_\times$. In Figure \ref{fig:Sq_vs_t}(b) we show that the scaling
\begin{equation}
S(q,t)=q^{-(1+2\alpha)} G(q t^{1/z}),
\label{eq:Sfamilyviscek} 
\end{equation}
with $\alpha=0.42 \pm 0.1$ and $z=1.6\pm 0.1$ and $G(x)\sim {\rm const}$ for large $x$ and $G(x)\sim x^{1+2\alpha}$ for small $x$ describes well the data for $q<0.05$.

\begin{figure}[h]
\includegraphics[width=\columnwidth]{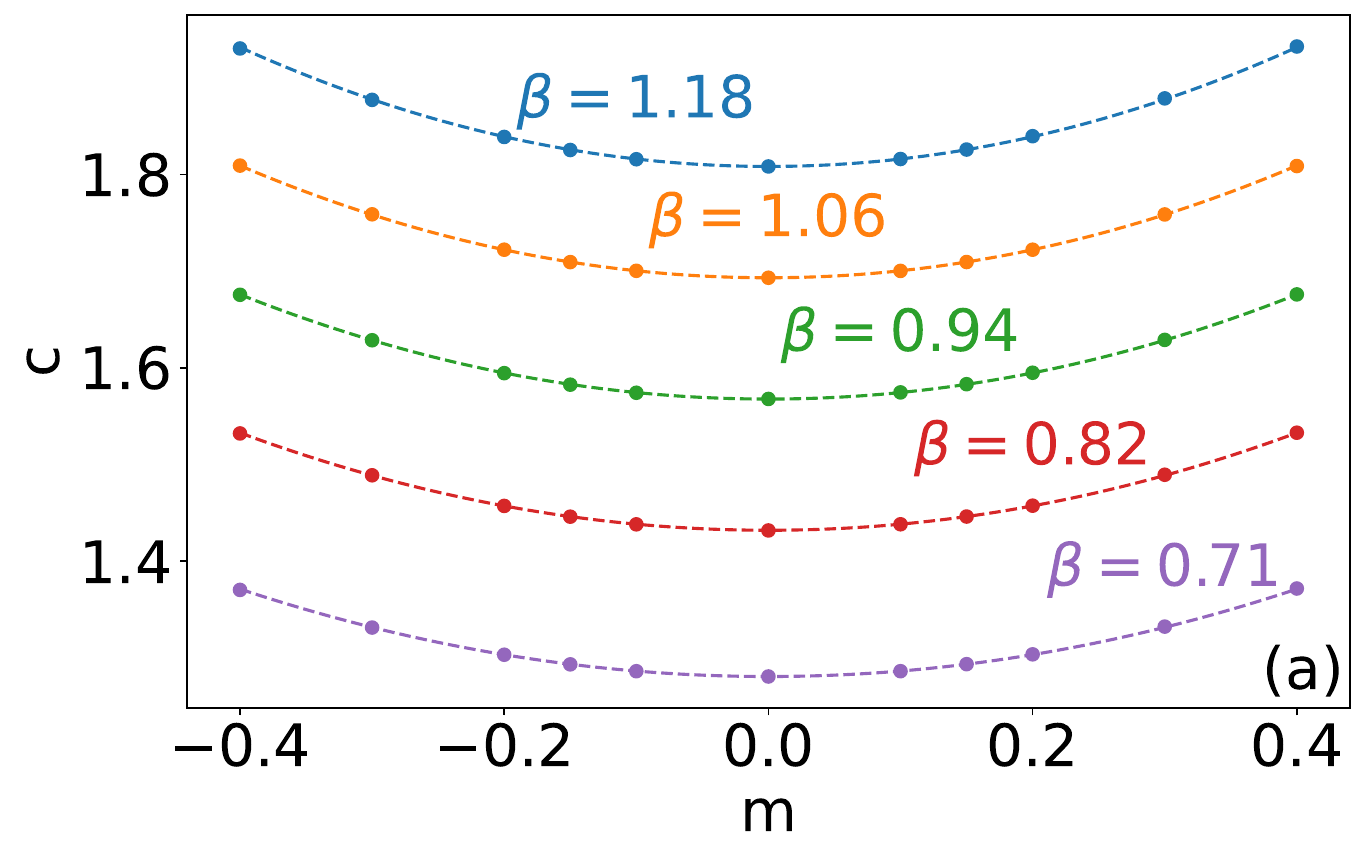}
    \includegraphics[width=\columnwidth]{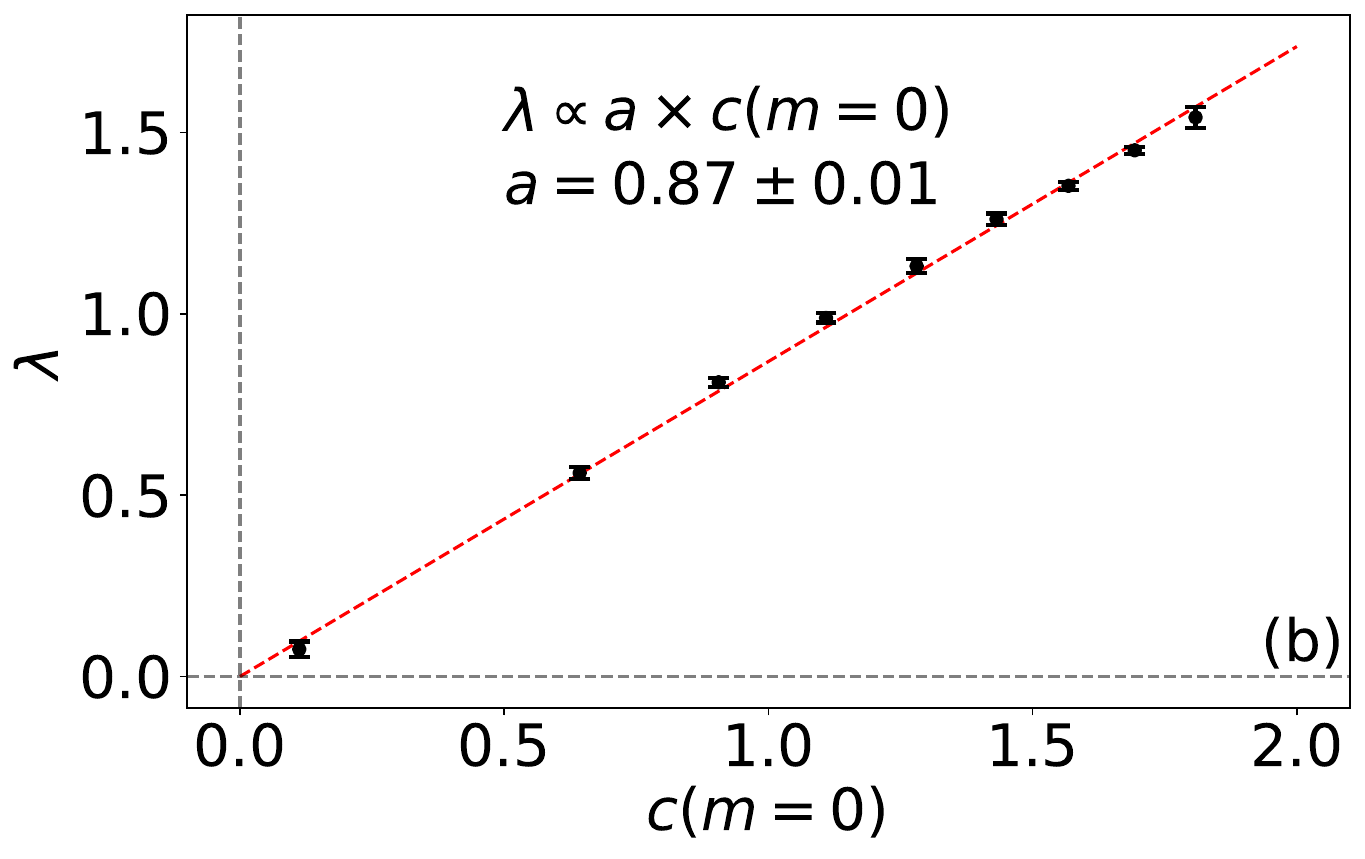}
    \caption{{(a)} Front velocity $c$ for tilted boundary conditions parametrized by $m$ and different values of $\beta$, in the RD case with $p=0.15$. 
    {(b)} KPZ parameter $\lambda$ as a function of the front velocity. The dashed line is the indicated linear fit.
    }
    \label{fig:lambda_vs_c}
\end{figure}

The exponents $\alpha=0.42 \pm 0.10$ and $z=1.6 \pm 0.10$, obtained for RD, SRD, and CD types of disorder at finite $c\sim {\mathcal{O}(1)}$ are not only robust but very close to the exact exponents $\alpha=1/2$ and $z=3/2$ of the one-dimensional KPZ equation describing the front position
\begin{align}
    \partial_t u(y,t) = f_0 +
    \nu \partial_y^2 u(y,t) + \frac{\lambda}{2} [\partial_y u(y,t)]^2 + \xi(y,t),
    \label{eq:kpzeq}
\end{align}
with $f_0$ the average velocity of the completely flat interface, $\nu>0$ an effective elastic constant and $\xi(y,t)$ an spatially and temporally uncorrelated noise satisfying 
\begin{eqnarray}
\langle{\xi(y,t)}\rangle 
&=& 0, \nonumber \\ 
\langle{\xi(y,t) \xi(y',t')}\rangle &=& 2 A \delta(y-y')\delta(t-t'),    
\label{eq:whitenoise}
\end{eqnarray}
with a noise amplitude $A>0$ over the noise ensemble.

In order to test the plausibility that the roughening of infection fronts in random-transmission media belongs to the one dimensional KPZ universality class, we first estimate the effective $\lambda$. From Eq.~\eqref{eq:kpzeq} we use that, for large enough $L_{\rm y}$ and periodic boundary conditions in the $y$-direction, the mean velocity of the interface should be
\begin{align}
c = {\int_y \frac{\langle\partial_t u(y,t)\rangle}{L_{\rm y}}}  \approx 
f_0 + \frac{\lambda}{2 L_{\rm y}}
\int_0^{L_{\rm y}} {dy}\;
\langle{[\partial_y u(y,t)]^2}\rangle.
\end{align}
If we now use Less-Edwards boundary conditions in the $y$-direction to impose an average tilt on the interface such that 
$\langle{\partial_y u}\rangle = m$, we get that $c(m) = c(m=0)+\frac{\lambda}{2}m^2$ and thus a method to obtain $\lambda$ from fitting $c(m)$ vs $m$~\cite{BarabasiBook}.
Applying this method to the infection front crest $u(y,t)$ for different $\overline{\beta}=\beta(1-p)$ 
in the RD case, 
corresponding to different velocities $c(m=0)$, we get the results of Figure \ref{fig:lambda_vs_c}(a). 
Curves for different velocities allow to fit $\lambda$ accurately  as a function of $c(m=0)$. 
In Figure \ref{fig:lambda_vs_c}(b) we show that $\lambda \approx a \times c(m=0)$ with $a=0.87 \pm 0.05$, 
with an ordinate that can be fairly neglected, as its value $\sim 0.01$ is always within its error bar. 
Equivalent results are found for the SRD and CD cases within their error bars. 
These results show that there is a positive effective KPZ parameter $\lambda$ directly proportional 
to the velocity in the absence of tilt $c(m=0)$, 
and that $\lambda \to 0$ as $c(m=0)\to 0$. 
Therefore, the effective interface description of the infection front has a KPZ term 
which is kinetically induced. 
The existence of a kinetically induced KPZ term is somehow expected 
for interfaces growing along their local normals in a statistically isotropic medium, 
though in the ideal case we would expect $\lambda=a \times c(m=0)$, with $a=1$, 
closer but slightly larger than the $\lambda=0.87\pm 0.05$ we find.

The plausibility of an effective noise $\xi(y,t)$ within the infection front finds support in the fact that as the front progresses, it encounters regions with different and uncorrelated quenched transmission rates. Consequently, in the co-moving frame, the front experiences temporally and spatially fluctuating forces, leading to fluctuating local velocities. Given that the front moves at a finite average velocity and the spatial disorder correlation is short-ranged in all directions, it is anticipated that the temporal correlation of the effective noise will also be short-range and uncorrelated along the front. 
This approximation aligns with the minimalist model outlined in Eq.~\eqref{eq:kpzeq} and in section 
\ref{sec:effectivenoise} we provide an approximate analytical argument showing how the effective noise on the moving front arises from quenched disorder.  
Furthermore, the anticipation is that an effective elasticity is present. This assumption is supported by the fact that if $\nu=0$ in Eq.~\eqref{eq:kpzeq}, the anticipated values would be $\alpha=1/3$ and $z=5/3$, which differ from the values we have empirically found: $\alpha=0.42 \pm 0.10$ and $z=1.6 \pm 0.10$.

It is important to acknowledge that despite the aforementioned arguments, there still exist slight disparities between the precise exponents $\alpha=1/2$ and $z=3/2$, and the estimated average exponents we obtained, $\alpha=0.42 \pm 0.10$ and $z=1.6 \pm 0.10$. Pinpointing the exact origin of these differences is challenging; however, it is highly probable that they arise from corrections to scaling due to finite time and size effects. Notably, even simpler models like the elastic string in random media can exhibit such corrections. Identifying these corrections requires simulating extensive systems, involving a scale of approximately $\sim 2^{25}$ points to describe the interface in models with continuous displacements. 
Unfortunately, such scales are impractical for our current approach in simulating the two-dimensional SIR model of infection spreading. 
Scaling corrections are nevertheless anticipated to be larger compared to those affecting thin interfaces with pure white noise (as in Eq. \eqref{eq:whitenoise}) due to the finite intrinsic characteristic front width $\Delta \sim D/c$ \cite{MurrayIIBook,Kolton2019}, as this length expands both the range of spatial and temporal correlations of the effective noise acting on the front, $\sim \Delta$ and $\sim \Delta/c$ respectively. 
These short-range correlations, although not impacting the convergence to the KPZ  exponent, may introduce a potentially extensive crossover~\cite{Mathey2017} and effective exponents. Interestingly, the corrections of $\alpha$ are expected to be negative, while for $z$ are expected to be positive~\cite{Lam1992} with respect to the universal asymptotic values. This heuristic may explain why $z=1.6\pm 0.10$ exceeds the exact KPZ value $z=3/2$, and why $\alpha=0.42 \pm 0.10$ falls below the exact KPZ value $\alpha=1/2$. Notably, when considering that $\Delta \approx D/c$, it further clarifies why in proximity to the threshold where $c$ is diminutive and consequently $\Delta$ is substantial, 
the effective roughness exponent descends even lower, approximately $\alpha \approx 0.3$, as noted in Ref.\cite{Kolton2019}. This rationale lends credibility to the expectation that the one-dimensional KPZ belongs to the same universality class. Interestingly, 
ecological invasion models \cite{OMalley2009} and flame fronts~\cite{Provatas1995}, were also shown to belong to the same universality class.

In summary, from the results of the present section we conclude:
\begin{itemize}
    \item The kinetic roughening of infection fronts is universal, with the one-dimensional KPZ the most probable universality class.
    \item A kinetically induced KPZ term and an effective short-range correlated dynamical noise are present.
\end{itemize}

\section{Conclusions}

Our investigation focused on analyzing infection fronts within the spatial 
SIR model on two-dimensional random media through numerical simulations. 

Our findings reveal non-universal behaviors in the velocity and harmfulness of these fronts, 
showcasing their sensitivity to the specific details of disorder. Notably, 
our study demonstrates that a heterogeneous distribution of a fixed ``vaccination amount'' or fixed ``protection resources'', 
aimed at locally reducing the transmission-rate  within a uniformly distributed susceptible population, 
outperforms the homogeneous distribution. 

Contrarily to the later conclusions, the large-scale structure of the advancing 
front crest appears to follow a 
universal robust pattern, closely aligning with the one-dimensional KPZ universality class, 
consistent with our finding of a kinetically generated KPZ term and dynamic noise induced by 
the front sliding over quenched disorder.

\section{Acknowledgments}
We thank M. M. Denham, E. E. Ferrero, E. A. Jagla and G. Abramson for enlightening  discussions. A. B. K. and K. L. are members of Consejo
Nacional de Investigaciones Científicas y Técnicas, Argentina (CONICET). We acknowledge Agencia Nacional de Promoción Científica y Tecnológica, Argentina (PICT 2019-03558 and PICT 2019-1991). 

\appendix
\section{Numerical Implementation} 
\label{sec:numerics}

Equations \eqref{eq:Seq}-\eqref{eq:Ieq} are numerically solved using a straightforward finite difference Euler explicit scheme in a square grid,
\begin{eqnarray}
I^{n+1}_{ij} &= I^n_{ij}
+ \delta {\tilde t} \left[\tilde{\beta}_{ij} S^n_{ij}I^n_{ij} - \tilde{\gamma} I^n_{ij} \right] 
+ 
\nonumber \\
\delta{\tilde t} \;\tilde{D} &( I^n_{i+1,j} + I^n_{i-1,j}
+ I^n_{i,j+1}+I^n_{i,j-1} 
-4 I^n_{ij}) \nonumber \\
S^{n+1}_{ij} &= S^{n}_{ij} - \delta {\tilde t}\; \tilde{\beta}_{ij} S^n_{ij} I^n_{ij}.
\label{eq:Ieqdiscretized}
\end{eqnarray}
where the sub-indices denote the discretized two-dimensional space coordinates, and the super-index the discretized time variable.
By measuring time in units of $\beta^{-1}$ and space in units of $\sqrt{D/\beta}$ we have the dimensionless parameters $\delta{\tilde t}=\beta \delta t$, and $\tilde{\gamma}=\gamma/\beta$, $\tilde{D}=D  /(\beta\delta x^2)$, where $\delta x$ is the spatial discretization in both directions and $\delta t$ the time discretization. 
An RD type of disorder is implemented by assigning $\tilde{\beta}_{ij}=0$ with probability $p$ or $\tilde{\beta}_{ij}=1$ with probability $1-p$. An SRD type of disorder is obtained by ${\tilde \beta}^{\rm SRD}_{ij} = 
({\tilde \beta}^{\rm RD}_{(i+1)j}+{\tilde \beta}_{(i-1)j}^{\rm RD}+{\tilde \beta}_{i(j+1)}^{\rm RD}+{\tilde \beta}^{\rm RD}_{i(j-1)})/4
$, and the DC is obtained by ${\tilde \beta}^{\rm CD}_{ij} 
\propto \Theta({\tilde \beta}^{\rm SRD}_{ij}-x)$, with $\Theta$ the Heaviside function where $x$ is chosen so to tune $\overline{\tilde \beta}$.

To iterate Eqs.\eqref{eq:Ieqdiscretized} we developed a parallel code based on CUDA \cite{nickolls2008scalable} in the Python language through the CuPy \cite{nishino2017cupy} package. This implementation is well suited for large systems $L_{\rm x} \times L_{\rm y} > 2^{20}$ which are necessary to study the kinetic roughening problem with a  proper precision. The code used in this work is freely available in Ref.~\onlinecite{CupyCode}.

\section{Effective noise}
\label{sec:effectivenoise}
We provide an approximate analytical argument demonstrating how the effective noise on the moving front arises from quenched disorder.
If the coarse-grained description of the front is governed by a KPZ equation we expect, by integrating Eq.~\eqref{eq:kpzeq} over $y$, that 
\begin{align}
\partial_t u_{\rm cm}(t) \approx c + \frac{1}{L_{\rm y}}\int_y \xi(y,t) \approx c + \delta c(t),
\label{eq:velfluckpz}
\end{align}
with $\delta c(t)$ an uncorrelated noise such that $
\int_t \langle \delta c(t)\delta c(t') \rangle \approx A/L_y$.
On the other hand, since we have $c =  G(\overline{\beta})$ for the actual front, with $G$ describing the mean steady-state velocity as a function of $\overline{\beta}$ for each type of disorder, we can approximate the instantaneous center of mass velocity by
\begin{align}
    \partial_t u_{\rm cm}(t) &\approx G\left(
    \overline{\beta}+
    \frac{1}{2L_{\rm y}\Delta}
    \int_{x=u_{\rm cm}(t)-\Delta}^{x=u_{\rm cm}(t)+\Delta} \int_y (\beta_{x,y}-\overline{\beta})  \right) \nonumber \\
    % &\approx
    % G\left(
    % \overline{\beta}+
    % \frac{1}{2L_{\rm y}\Delta}
    % \int_{x=c t -\Delta}^{x=c t+\Delta} \int_y (\beta_{x,y}-\overline{\beta})  \right) \nonumber \\
    &\approx
    c + 
    \frac{1}{L_{\rm y}}
    \int_y \left[ \frac{G'(\overline{\beta})}{2\Delta}
    \int_{x=-\Delta}^{x=\Delta} (\beta_{ct+x,y}-\overline{\beta}) \right]
    \nonumber \\
    &= c + \delta c(t).
    \label{eq:velfluc}
\end{align}
In the second term of Eq.~\eqref{eq:velfluc}, we used that the instantaneous center of mass velocity is approximately given by the function $G$ evaluated at the average of $\beta_{\bf r}$ over a region of size $\Delta$ around $u_{\rm cm}(t)$, with $\Delta$ being the front width (i.e. a moving window average). 
In the third term, we expanded to first order in the fluctuations around $\overline{\beta}$ and made the approximation $u_{\rm cm}\approx c t$ in the integrand.
In the fourth term we identify the center of mass velocity fluctuation term $\delta c(t)$ around the steady-state velocity $c=G(\overline{\beta})$. Comparing Eqs.\eqref{eq:velfluckpz} and \eqref{eq:velfluc} we can see that the effective noise is 
\begin{align}
    \xi(y,t) \approx
    \frac{G'(\overline{\beta})}{2\Delta}
    \int_{x=-\Delta}^{x=\Delta} (\beta_{ct+x,y}-\overline{\beta}).
\end{align}
Using that $\beta_{\bf r}$ is short-range uncorrelated in both space directions, computing $\langle \xi(y,t)\xi(y',t')\rangle$ it is easy to see that $\xi(y,t)$ is  completely uncorrelated along the $y$-direction, and  short-range correlated in time with a finite range $\tau \sim \Delta/c$. 
The intensity of the effective noise is then $A \propto G'^2(\overline{\beta})
\langle [\beta_{\bf r}-\overline{\beta}]^2 \rangle /\Delta^2$. As expected, the noise increases with decreasing the front width $\Delta$ or with increasing the fluctuations of $\beta_{\bf r}$. Interestingly, the noise also depends on $\overline{\beta}$, through $G'(\overline{\beta})$. 

\bibliography{biblio}{}
\bibliographystyle{plain}
\bibliographystyle{unsrt}

\end{document}